\documentclass{PoS}

\def\lsim{\mathrel{\rlap{\lower4pt\hbox{\hskip1pt$\sim$}}
    \raise1pt\hbox{$<$}}}         
\def\gsim{\mathrel{\rlap{\lower4pt\hbox{\hskip1pt$\sim$}}
    \raise1pt\hbox{$>$}}}         

\title{Production of two $c \bar c$ pairs \\
in double-parton scattering within $k_t$-factorization}

\ShortTitle{Production of two $c \bar c$ pairs}

\author{\speaker{Antoni SZCZUREK}
\thanks{This work was supported by the Polish National Science Centre 
(on the basis of decision No.~DEC-2011/01/B/ST2/04535).}\\
University of Rzesz\'ow, PL-35-959 Rzesz\'ow, Poland, and\\
Institute of Nuclear Physics PAN, PL-31-342 Cracow, Poland\\
E-mail: \email{Antoni.Szczurek@ifj.edu.pl}}

\author{Rafa{\l} MACIU{\L}A\\
Institute of Nuclear Physics PAN, PL-31-342 Cracow, Poland\\
E-mail:\email{Rafal.Maciula@ifj.edu.pl}}

\abstract{
We discuss production of two pairs of $c \bar c$ in proton-proton
collisions at the LHC.
Both double-parton scattering (DPS) and single-parton scattering (SPS)
contributions are included in the analysis. Each step of DPS is
calculated within $k_t$-factorization approach.
The conditions how to identify the DPS contribution are presented. 
The discussed mechanism leads to the production of pairs of mesons:
each containing $c$ quarks or each containing $\bar c$ antiquarks. 
We discuss corresponding production rates and 
some differential distributions for $(D^0 D^0$ + $\bar D^0 \bar D^0)$ 
production. Within large theoretical uncertainties the predicted DPS
cross section is fairly similar to the cross section measured recently
by the LHCb collaboration.
The best description is obtained with the Kimber-Martin-Ryskin (KMR)
unintegrated gluon distribution. The contribution of SPS,
calculated in the high-energy approximation, turned out to be rather
small. Finally, we emphasize significant contribution of DPS mechanism 
to inclusive charmed meson spectra measured recently by ALICE, ATLAS and LHCb.
}

\FullConference{XXI International Workshop on Deep-Inelastic Scattering
  and Related Subject -DIS2013,\\
		22-26 April 2013\\
		Marseilles,France}

\begin{document}

\section{Introduction}

There is recently growing interest in studying double-parton
scattering (DPS) effects 
(see e.g. \cite{Bartalini2011} and references therein).
Recently we have shown that the production of $c \bar c c \bar c$
is an ideal place to study DPS effects \cite{LMS2012}.
Here, the quark mass is small enough to assure that the cross section 
for DPS is very large, and large enough to treat the problem within
pQCD. The calculation performed in Ref.~\cite{LMS2012} was done in 
the leading-order (LO) collinear approximation. Recently \cite{MS2013} 
we have shown that this is not sufficient when comparing the results of 
the calculation with real experimental data. In the meantime the LHCb 
collaboration presented new experimental data for simultaneous
production of two charmed mesons \cite{LHCb-DPS-2012}. 
In spite of limited acceptance they have observed large percentage of 
events with two mesons, both containing $c$ quark.

In our recent analysis \cite{MS2013} we have argued that
the LHCb data provide a footprint of double parton scattering.
In addition, we have also estimated $c \bar c c \bar c$ production via 
single-parton scattering (SPS) within a high-energy approximation 
\cite{SS2012}. This approach seems to be an efficient tool especially
when the distance in rapidity between $c c$ or/and $\bar c \bar c$ is large.

Another piece of evidence for the DPS effects is that their absence leads to a
missing contribution to inclusive charmed meson production, as noted
in Ref.~\cite{MS2013-charmed-mesons}.
The measured inclusive cross sections include events where two 
$D$ (or two $\bar D$) mesons are produced, therefore corresponding theoretical predictions should also be corrected for the DPS effects.

In Ref.~\cite{Berezhnoy2012} the authors estimated DPS contribution
based on the experimental inclusive $D$ meson spectra measured at LHC.
In their approach fragmentation was included only in terms
of the branching fractions for the transition $c \to D$. In our approach
we include full kinematics of hadronization process.

\section{Sketch of the formalism}

\begin{figure}[!h]
\begin{center}
\includegraphics[width=4cm]{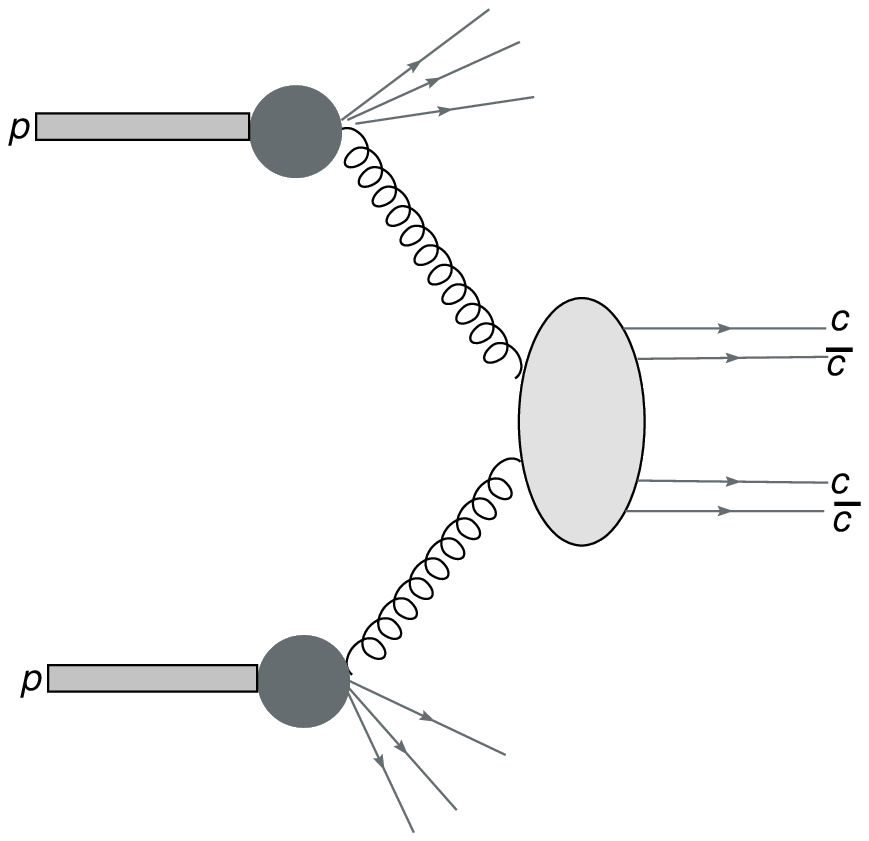}
\includegraphics[width=4cm]{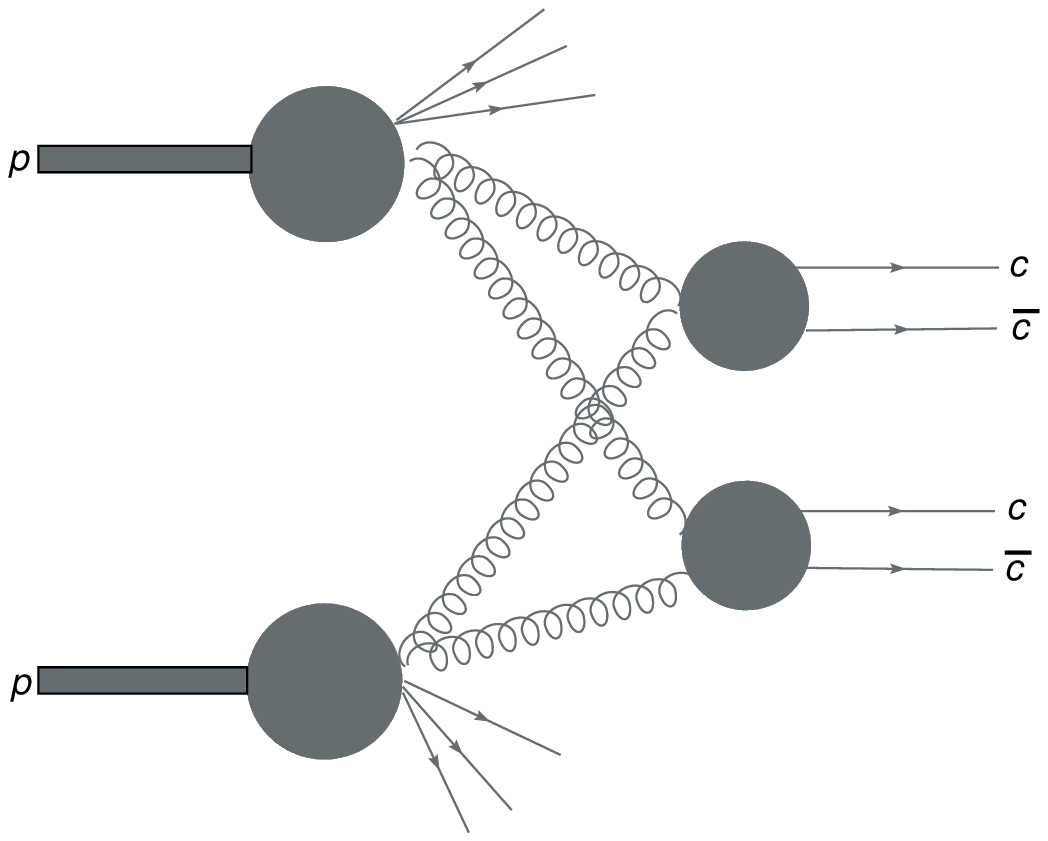}
\end{center}
   \caption{
\small SPS (left) and DPS (right) mechanisms of $(c \bar c) (c \bar c)$ 
production.  
}
\label{fig:mechanisms_ccbarccbar}
\end{figure}

Two possible mechanisms of the production of two $c \bar c$ pairs are shown in
Fig.\ref{fig:mechanisms_ccbarccbar}.

The cross section for differential distribution in a simple
double-parton scattering in leading-order collinear approximation 
can be written as
\begin{equation}
\frac{d \sigma}{d y_1 d y_2 d^2 p_{1t} d y_3 d y_4 d^2 p_{2t}}  \\ =
\frac{1}{ 2 \sigma_{eff} }
\frac{ d \sigma } {d y_1 d y_2 d^2 p_{1t}} \cdot
\frac{ d \sigma } {d y_3 d y_4 d^2 p_{2t}} 
\label{differential_distribution}
\end{equation}
which by construction reproduces the formula for integrated cross
section \cite{LMS2012}.
This cross section is formally differential in 8 dimensions but can be 
easily reduced to 7 dimensions noting that physics of unpolarized
scattering cannot depend on azimuthal angle of the pair or on azimuthal
angle of one of the produced $c$ ($\bar c$) quark (antiquark).
This can be easily generalized by including QCD evolution effects
\cite{LMS2012}.

In the $k_t$-factorization approach the differential cross section for
DPS production of $c \bar c c \bar c$ system, assuming factorization of 
the DPS model, can be written as: 
\begin{eqnarray}
\frac{d \sigma^{DPS}(p p \to c \bar c c \bar c X)}{d y_1 d y_2 d^2 p_{1,t} d^2 p_{2,t} 
d y_3 d y_4 d^2 p_{3,t} d^2 p_{4,t}} = \nonumber \;\;\;\;\;\;\;\;\;\;\;\;\;\;\;\;\;\;\;\;\;\;\;\;\;\;\;\;\;\;\;\;\;\;\;\;\;\;\;\;\;\;\;\;\;\;\;\;\;\;\;\;\;\;\;\;\;\;\;\;\;\;\;\;\;\;\; \\ 
\frac{1}{2 \sigma_{eff}} \cdot
\frac{d \sigma^{SPS}(p p \to c \bar c X_1)}{d y_1 d y_2 d^2 p_{1,t} d^2 p_{2,t}}
\cdot
\frac{d \sigma^{SPS}(p p \to c \bar c X_2)}{d y_3 d y_4 d^2 p_{3,t} d^2 p_{4,t}}.
\end{eqnarray}
These formulae assume that the two parton subprocesses are not correlated one
with each other. 
The parameter $\sigma_{eff}$ in the denominator of above formulae can 
be understood in the impact parameter space as:
\begin{equation}
\sigma_{eff} = \left[ \int d^{2}b \; (T(\vec{b}))^{2} \right]^{-1},
\end{equation}
where the overlap function
\begin{equation}
T ( \vec{b} ) = \int f( \vec{b}_{1} ) f(\vec{b}_{1} - \vec{b} ) \; d^2 b_{1},
\end{equation}
In the presented here analysis cross section for each step is calculated
in the $k_t$-factorization approach:
\begin{eqnarray}
\frac{d \sigma^{SPS}(p p \to c \bar c X_1)}{d y_1 d y_2 d^2 p_{1,t} d^2 p_{2,t}} 
&& = \frac{1}{16 \pi^2 {\hat s}^2} \int \frac{d^2 k_{1t}}{\pi} \frac{d^2 k_{2t}}{\pi} \overline{|{\cal M}_{g^{*} g^{*} \rightarrow c \bar{c}}|^2} \nonumber \\
&& \times \;\; \delta^2 \left( \vec{k}_{1t} + \vec{k}_{2t} - \vec{p}_{1t} - \vec{p}_{2t}
\right)
{\cal F}(x_1,k_{1t}^2,\mu^2) {\cal F}(x_2,k_{2t}^2,\mu^2),
\nonumber
\end{eqnarray}
\begin{eqnarray}
\frac{d \sigma^{SPS}(p p \to c \bar c X_2)}{d y_3 d y_4 d^2 p_{3,t} d^2 p_{4,t}} 
&& = \frac{1}{16 \pi^2 {\hat s}^2} \int \frac{d^2 k_{3t}}{\pi} \frac{d^2 k_{4t}}{\pi} \overline{|{\cal M}_{g^{*} g^{*} \rightarrow c \bar{c}}|^2} \nonumber \\
&&\times \;\; \delta^2 \left( \vec{k}_{3t} + \vec{k}_{4t} - \vec{p}_{3t} - \vec{p}_{4t}
\right)
{\cal F}(x_3,k_{3t}^2,\mu^2) {\cal F}(x_4,k_{4t}^2,\mu^2).
\end{eqnarray}
The matrix elements for $g^* g^* \to c \bar c$ (off-shell gluons) are 
calculated including transverse momenta of initial gluons
\cite{CCH91,CE91,BE01}.
The unintegrated ($k_t$-dependent) gluon distributions (UGDFs) in the
proton are taken from the literature \cite{KMR,KMS,Jung}.

How the single scattering contribution is calculated is explained in
Ref.\cite{MS2013-charmed-mesons}. In this approach a high-energy approximation
is used, and the calculation should be reliable for larger
rapidity distance between two $c c$ or two $\bar c \bar c$ quarks.
A full calculation of the single scattering contribution is in
progress \cite{HS2013}.

\section{Results}
\label{section:Results}

\begin{figure}[!h]
\begin{center}
\includegraphics[width=7.0cm]{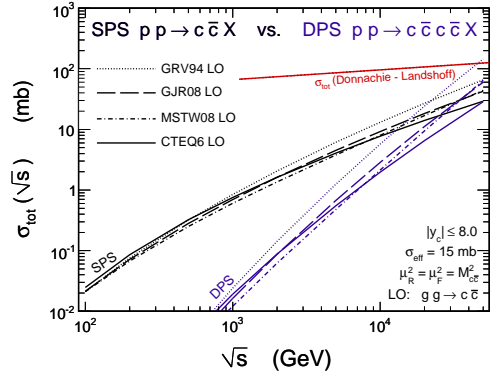}
\end{center}
   \caption{
\small Total LO cross section for single
$c \bar c$ pair and SPS and DPS $c \bar c c \bar c$ production
as a function of center-of-mass energy.  
}
 \label{fig:single_vs_double_LO}
\end{figure}

In Fig.~\ref{fig:single_vs_double_LO} we
compare cross sections for the single $c \bar c$ pair production as well
as for single-parton and double-parton scattering $c \bar c c \bar c$
production as a function of proton-proton center-of-mass energy. 
At low energies the conventional single $c \bar c$ pair production
cross section is much larger. 
The cross section for SPS production
of $c \bar c c \bar c$ system is more than two orders of magnitude smaller
than that for $c \bar c$ production. For reference we also show 
the parametrization of proton-proton total cross section as a function
of center-of-mass energy.
At higher energies the DPS contribution
of $c \bar c c \bar c$ quickly approaches that for single $c \bar c$ 
production as well as the total cross section.

\begin{figure}[!h]
\begin{center}
\includegraphics[width=7.0cm]{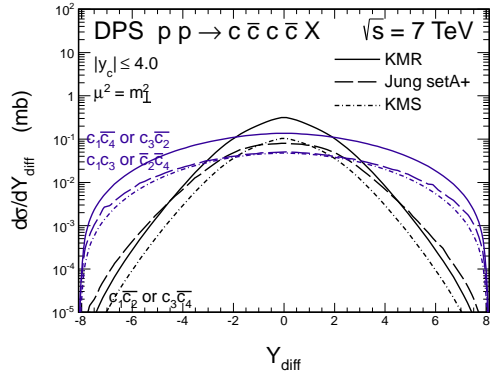}
\end{center}
   \caption{
\small Distribution in rapidity distance between quarks/antiquarks 
$Y_{diff}$ from the same ($c_1\bar c_2$ or $c_3\bar c_4$) and 
from different scatterings ($c_1\bar c_4$ or $c_3\bar c_2$ or $c_1 c_3$ 
or $\bar c_2\bar c_4$), calculated with different UGDFs.
}
\label{fig:correlations_kt_factorization_1}
\end{figure}

In Ref.~\cite{LMS2012} we have proposed several correlation distributions
to be studied in order to identify the DPS effects.
Here in Fig.~\ref{fig:correlations_kt_factorization_1} we show only 
distributions in rapidity difference of quarks/antiquarks 
$Y_{diff} = y_c - y_{\bar c}$ from the
same scattering ($c_1\bar c_2$ or $c_3\bar c_4$) and from different 
scatterings ($c_1\bar c_4$ or $c_3\bar c_2$ or $c_1 c_3$ or $\bar
c_2\bar c_4$) for various UGDFs. 
The shapes of distributions in the figure are almost identical as 
that obtained in LO collinear approach in Ref.~\cite{LMS2012}. 
One can clearly see that the double-scattering effects dominate
in the region of large rapidity distances. This is a potential place
to identify them. As discussed in Ref.\cite{MS2013} this is not very
easy option with the existing detectors. Further studies are needed.

\begin{figure}[!h]
\begin{minipage}{0.47\textwidth}
 \centerline{\includegraphics[width=1.0\textwidth]{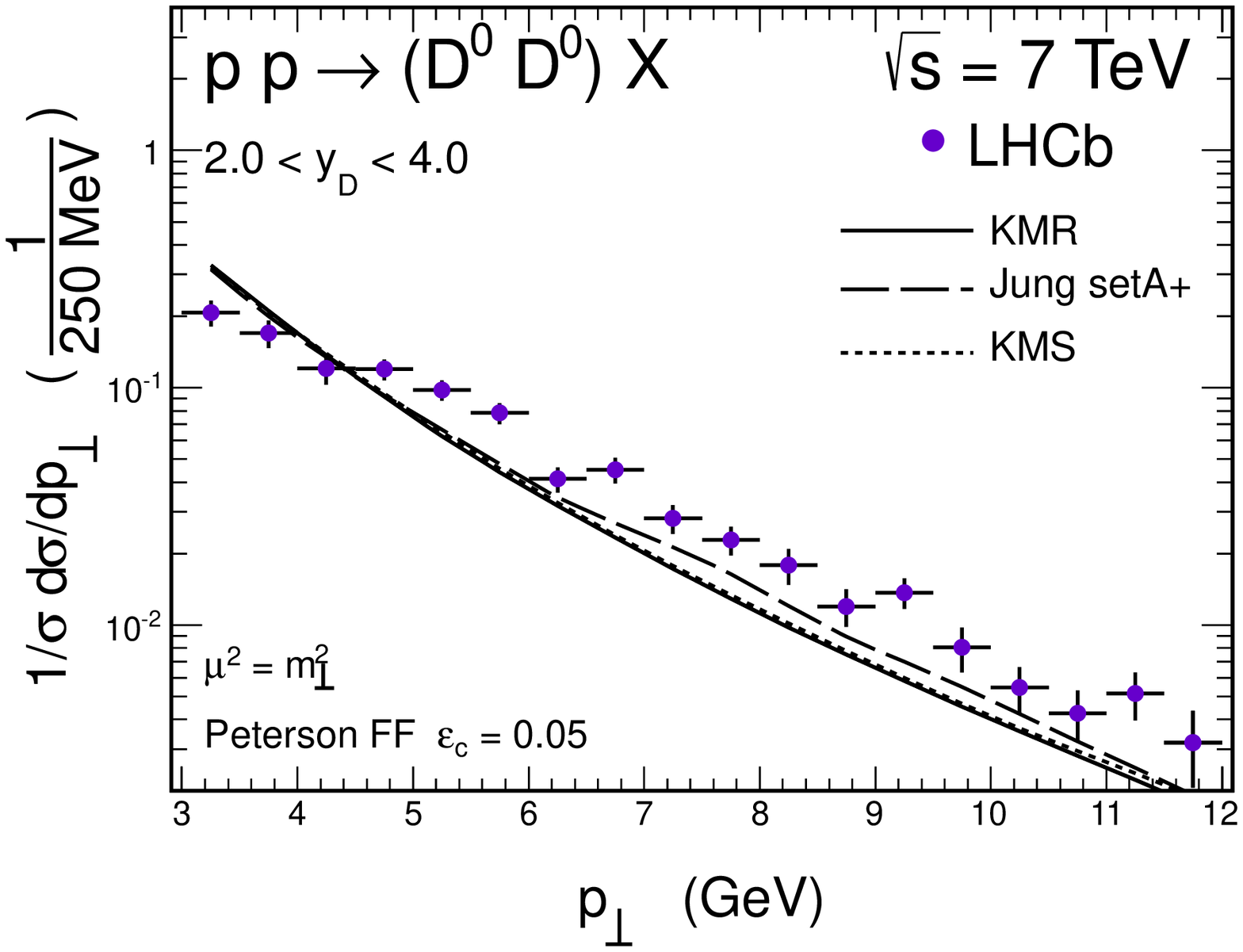}}
\end{minipage}
\hspace{0.5cm}
\begin{minipage}{0.47\textwidth}
 \centerline{\includegraphics[width=1.0\textwidth]{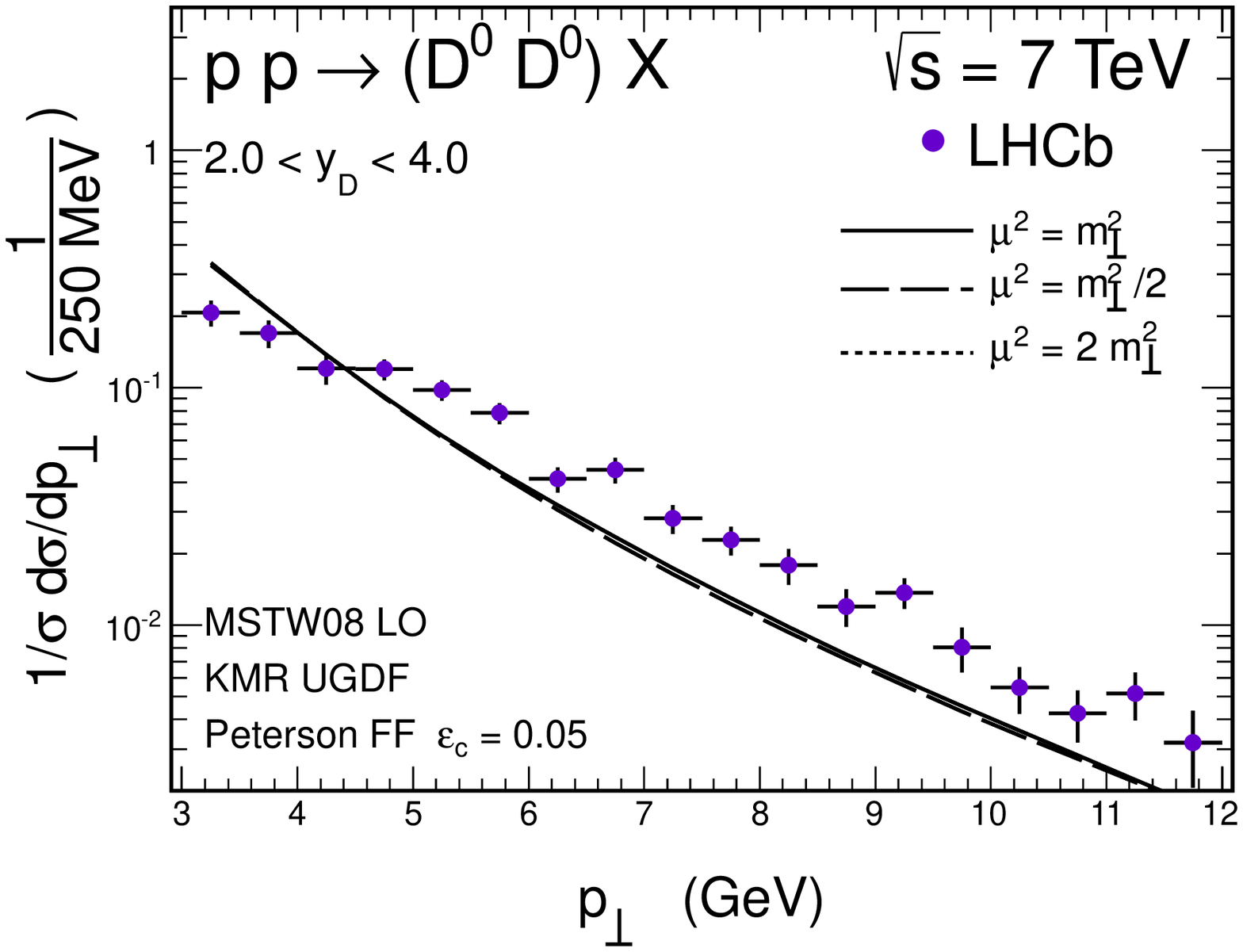}}
\end{minipage}
   \caption{
\small Transverse momentum distribution of $D^0$ mesons from the
$D^0 D^0$ pair contained in the LHCb kinematical region. The left panel shows
dependence on UGDFs, while the right panel illustrates dependence of the
result for the KMR UGDF on the factorization/renormalization scales. The MSTW08 collinear distribution
was used to generate KMR UGDF.
}
\label{fig:pt-lhcb-DD-1}
\end{figure}

In Fig.~\ref{fig:pt-lhcb-DD-1} we present distribution in transverse
momentum of one of the $D^0$ mesons, provided that both are measured
within the LHCb experiment coverage.
Our theoretical distributions have shapes in rough agreement with the
experimental data. The shapes of the distributions are almost identical for 
different UGDFs used in the calculations (left panel) and are
almost independent of the choice of scales in the case of the KMR model 
(right panel).

\begin{figure}[!h]
\begin{minipage}{0.47\textwidth}
 \centerline{\includegraphics[width=1.0\textwidth]{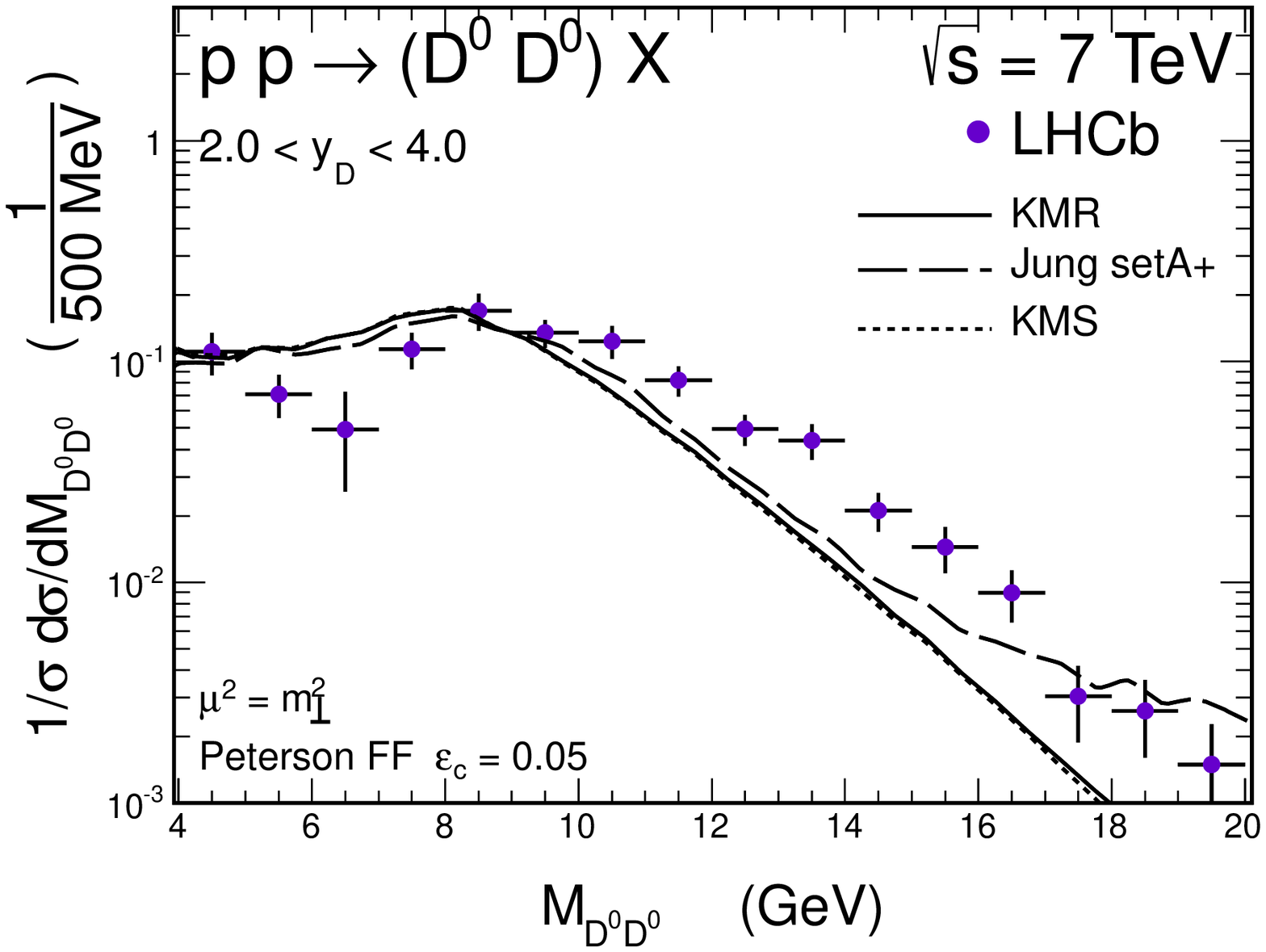}}
\end{minipage}
\hspace{0.5cm}
\begin{minipage}{0.47\textwidth}
 \centerline{\includegraphics[width=1.0\textwidth]{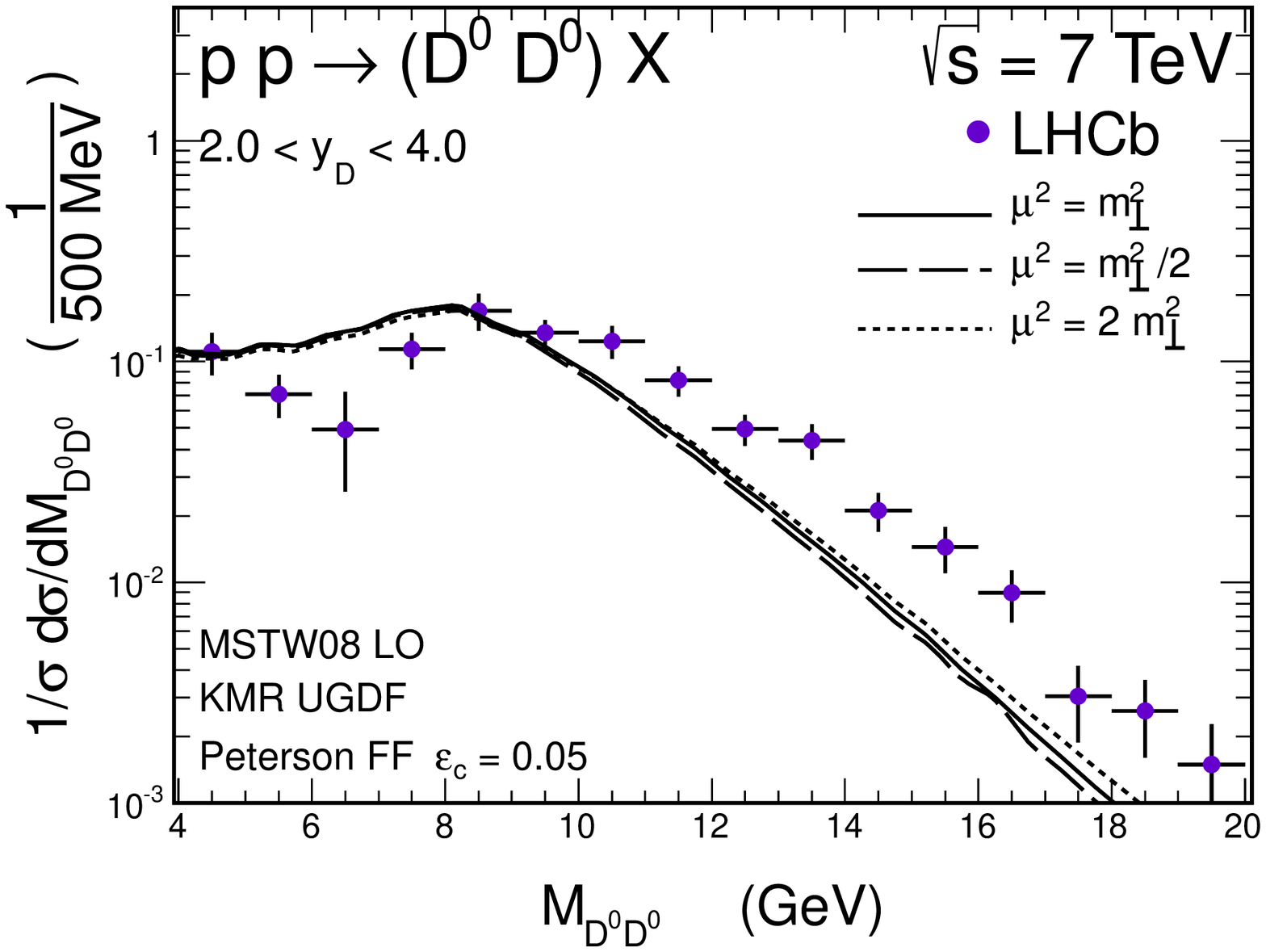}}
\end{minipage}
   \caption{
\small $M_{D^{0}D^{0}}$ invariant mass distribution for $D^0 D^0$ contained in the
LHCb kinematical region. The left panel shows
dependence on UGDFs, while the right panel illustrates dependence of the
result on the factorization/renormalization scales for KMR UGDF.
}
\label{fig:minv-lhcb-DD-2}
\end{figure}

In Fig.~\ref{fig:minv-lhcb-DD-2} we show distribution in the $D^0 D^0$
invariant mass $M_{D^{0}D^{0}}$ for both $D^0$'s measured in the kinematical region
covered by the LHCb experiment. Here the shapes of the distributions have the same behavior
for various UGDFs and are insensitive to changes of scales as in the previous figure.
The characteristic minimum at small invariant masses is a consequnce of
experimental cuts (see Ref.~\cite{MS2013-charmed-mesons}) and is rather well reproduced.
Our approach fails at large dimeson invariant masses. The large $M_{D^{0}D^{0}}$ invariant masses
are probably correlated to large scales $\mu_{1/2}^2$ and/or
$\mu_{3/4}^2$. If these can be related to the effects of factorization
violation discussed in ~\cite{Gustaffson2011} requires dedicated studies.  

\begin{figure}[!h]
\begin{minipage}{0.47\textwidth}
 \centerline{\includegraphics[width=1.0\textwidth]{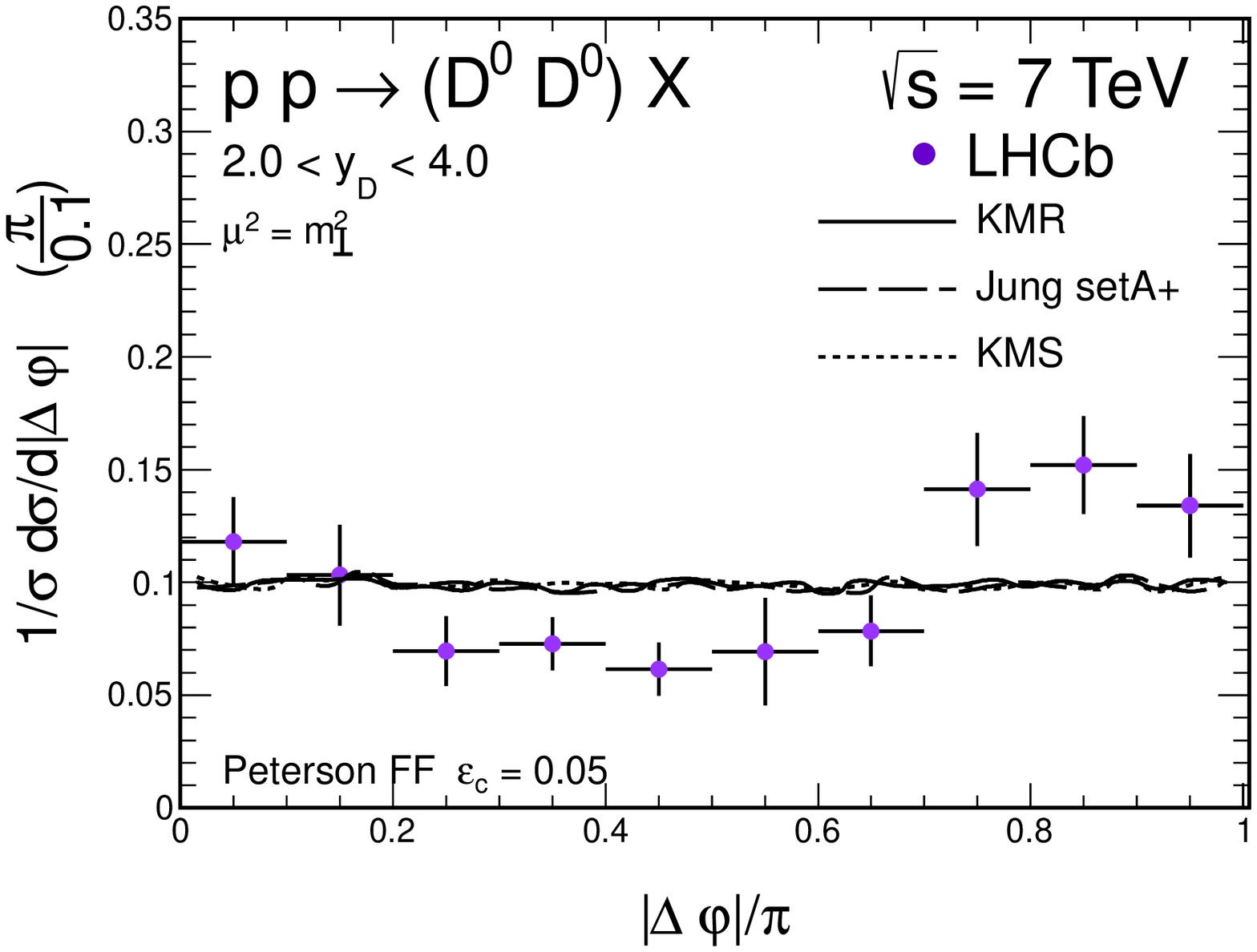}}
\end{minipage}
\hspace{0.5cm}
\begin{minipage}{0.47\textwidth}
 \centerline{\includegraphics[width=1.0\textwidth]{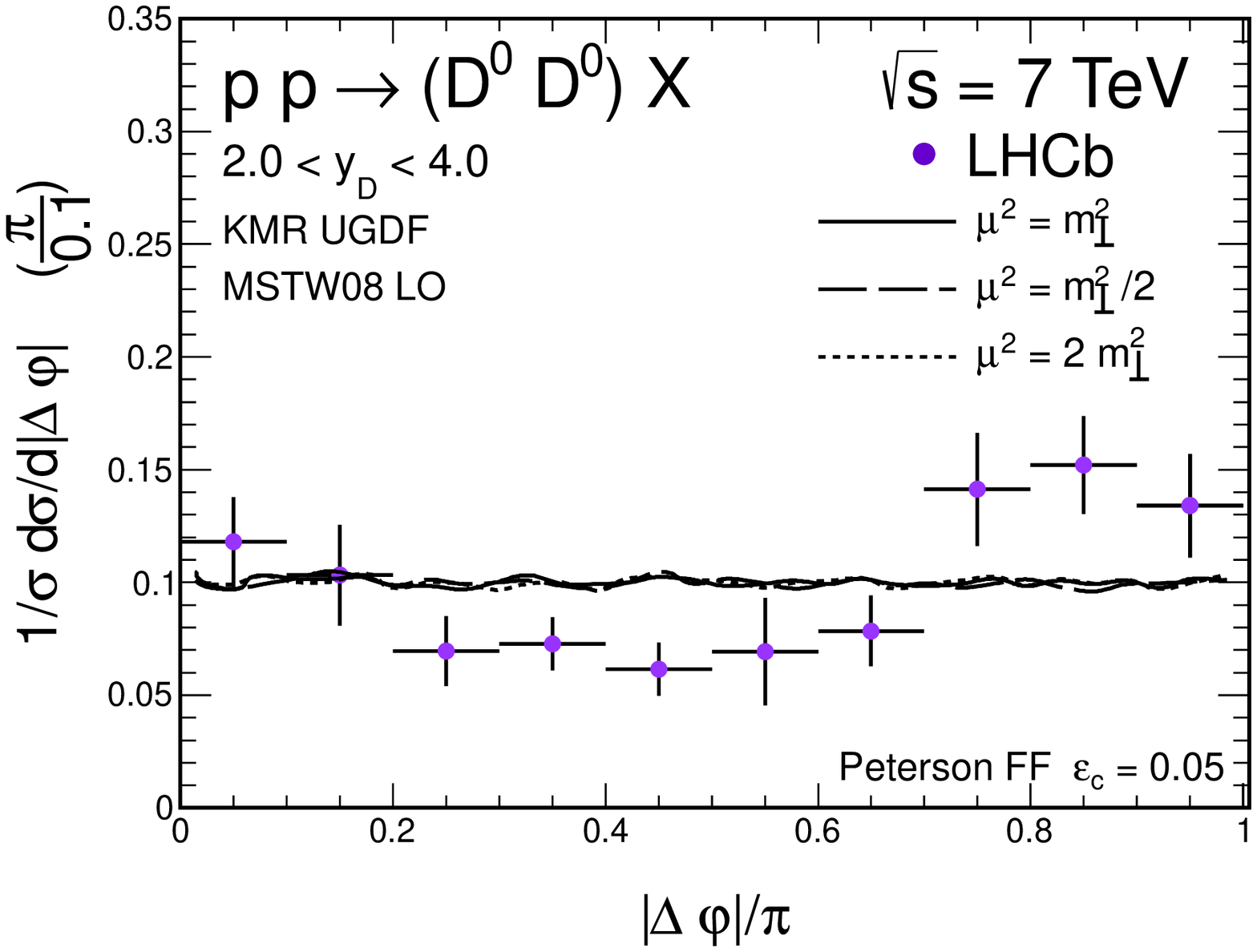}}
\end{minipage}
   \caption{
\small Distribution in azimuthal angle $\varphi_{D^{0}D^{0}}$ between both $D^0$'s.
The left panel shows
dependence on UGDFs, while the right panel illustrates dependence of the
result for the KMR UGDF on the factorization/renormalization scales.
}
\label{fig:phid-lhcb-DD-3}
\end{figure}

Finally in Fig.~\ref{fig:phid-lhcb-DD-3} we show distribution in
azimuthal angle $\varphi_{D^{0}D^{0}}$ between both $D^0$'s. While the
theoretical DPS contribution is independent of the relative azimuthal
angle, there is a small dependence on azimuthal angle in experimental distribution. 
This may show that there is some missing mechanism which gives
contributions both at small and large $\Delta \varphi$. However, this
discrepancy may be also an inherent property of the DPS factorized model
which does not allow for azimuthal correlations between particles
produced in different hard scatterings.
We wish to emphasize in this context that the angular azimuthal
correlation pattern for $D^{0}\bar{D^{0}}$, discussed in 
Ref.~\cite{MS2013-charmed-mesons}, and for $D^{0}D^{0}$
$(\bar{D^{0}}\bar{D^{0}})$, discussed here, are quite different. The
distribution for $D^{0}D^{0}$ $(\bar{D^{0}}\bar{D^{0}})$
is much more flat compared to the $D^{0}\bar{D^{0}}$ one which shows a
pronounced maximum at $\varphi_{D^{0}\bar{D^{0}}} = 180^{\circ}$ 
and $\varphi_{D^{0}\bar{D^{0}}} = 0^{\circ}$. This qualitative difference is
in our opinion a model independent proof of the dominance of DPS effects
in the production of $D^{0}D^{0}$ $(\bar{D^{0}}\bar{D^{0}})$.

\section{Conclusions}

In this presentation we have discussed production of $c \bar c c \bar c$
in the double-parton scattering (DPS) and single-parton scattering (SPS)
in the $g g \to c \bar c c \bar c$ subprocess. The double-parton
scattering is calculated in the factorized Ansatz with each step
calculated in the $k_t$-factorization approach, i.e. including
effectively higher-order QCD corrections.
 
The distribution in rapidity difference between
quarks/antiquarks from the same and different scatterings turned
out to have similar shape as in the LO collinear approach. The same
is true for invariant masses of pairs of quark-quark,
antiquark-antiquark and quark-antiquark, etc.
The distribution in transverse momentum of the pair from the same
scattering turned out to be similar to that for the pairs originating from
different scatterings. 

The total rates of the meson pair production depend on the unintegrated
gluon distributions. The best agreement with the LHCb result has been 
obtained for the Kimber-Martin-Ryskin UGDF. This approach, as discussed 
already in the literature, effectively includes higher-order QCD corrections.

As an example we have also calculated several differential distributions
for $D^0 D^0$ pair production. We have reproduced the main trends of the LHCb data
for transverse momentum distribution of $D^0$ $(\bar D^0)$ mesons
and $D^0 D^0$ invariant mass distribution. The distribution in azimuthal
angle between both $D^0$'s suggests that some mechanisms may be still
missing. The single parton scattering contribution, calculated in the
high energy approximation, turned out to be rather small. This is being
checked in exact $2 \to 4$ parton model calculations. 

The DPS mechanism of $c \bar c c \bar c$ production gives a new
significant contribution to inclusive charmed meson spectra \cite{MS2013}. 
For instance the description of the inclusive ATLAS, ALICE and LHCb data is
very difficult in terms of the conventional SPS ($c \bar c$) 
contribution \cite{MS2013-charmed-meson}. 

Summarizing, the present study of $c \bar c c \bar c$ reaction in the
$k_t$-factorization approach has shown that this reaction is
one of the best places for testing double-parton scattering effects.

\end{document}